\documentclass[%
reprint,
superscriptaddress,
amssymb,
aps,
prl,
floatfix,
]{revtex4-2}

\usepackage{graphicx}
\graphicspath{{figures/}}
\usepackage{rotating}
\usepackage{amsmath}
\usepackage{bbm}
\usepackage{subfigure}
\usepackage{color}
\usepackage{braket}
\usepackage{tikz}
\usepackage{hyperref}
\usepackage[normalem]{ulem}
\hypersetup{colorlinks, 
	linkcolor={blue!75!black!80!yellow},
	citecolor={blue!75!black!80!yellow}, 
	urlcolor={blue!75!black!80!yellow}
	}

\makeatletter \renewcommand\@make@capt@title[2]{%
\@ifx@empty\float@link{\@firstofone}{\expandafter\href\expandafter{\float@link}}%
\sffamily{\textbf{#1}}\@caption@fignum@sep#2 }% \makeatother

\begin{document}

\title{Modifying van der Waals Materials via Cavity Vacuum Fluctuations}

\author{Mohammad Hassan}
\affiliation{Department of Physics, City College of New York, New York, NY 10031, USA}
\affiliation{Department of Physics, The Graduate Center, City University of New York, New York, NY 10016, USA}
\author{Cankut Tasci}
\affiliation{Department of Physics, City College of New York, New York, NY 10031, USA}
\affiliation{Department of Physics, The Graduate Center, City University of New York, New York, NY 10016, USA}
\author{Leon Orlov-Sullivan}
\affiliation{Department of Physics, City College of New York, New York, NY 10031, USA}
\author{Leonardo A. Cunha}
\affiliation{Center for Computational Quantum Physics, Flatiron Institute, New York, 10010  NY,  USA}
\affiliation{Department of Chemistry and Biochemistry, Bates College, Lewiston, ME 04240, USA}
\author{Johannes Flick}
  \email[Electronic address:\;]{jflick@ccny.cuny.edu}
\affiliation{Department of Physics, City College of New York, New York, NY 10031, USA}
\affiliation{Department of Physics, The Graduate Center, City University of New York, New York, NY 10016, USA}
\affiliation{Center for Computational Quantum Physics, Flatiron Institute, New York, 10010  NY,  USA}

%%%%%%%%%%%%%%%%%%%%%%%%%%%%%%%%%%%%%%%%%%%%%%%%%%%%%%%%%%%%%%%%%%
%                            Abstract                            %
%%%%%%%%%%%%%%%%%%%%%%%%%%%%%%%%%%%%%%%%%%%%%%%%%%%%%%%%%%%%%%%%%%
\begin{abstract}
In the field of cavity quantum materials, vacuum fluctuations of optical cavities are used for modifying ground-state properties of quantum materials without external driving. Here, one example  is the van der Waals (vdW)/dispersion interaction in layered 2D vdW materials, where non-additive long-range correlations can dominate the interlayer binding. While cavity-induced changes of such interactions have been predicted and described using ab initio methods for molecular systems, no efficient description exists yet for extended materials. In this Letter, we close this gap by introducing a periodic formulation of the photon many-body dispersion (pMBD) functional within quantum electrodynamical density-functional theory (QEDFT). Applying this method with efficient $\textbf{q}$-point sampling to bilayer hBN and graphene, we predict cavity-modified stacking, increased equilibrium interlayer distances, and softened layer breathing modes with increasing light-matter coupling strength. Our results establish cavity vacuum fluctuations as a tuning knob for the structural properties of vdW materials.
\end{abstract}

\maketitle

\textit{Introduction:}
The embedding of quantum materials in optical cavities has emerged as a promising path towards modifying the ground-state properties of materials without external driving, using the vacuum fluctuations of the confined electromagnetic field \cite{reviewebbesen2021, Lu_cavityengineering, fluctuationengineeringbretscher2026}. Experimentally, cavity coupling has been shown to shift the metal-to-insulator transition \cite{jarc2023cavity}, to suppress \cite{cavityalteredsuperconductivityKeren2026} or enhance \cite{cavityenhancedsuperconductingresponseunderdopedmontanaro2026, cavityenhancedsuperconductivityzhang2026} superconductivity, to modify integer and fractional quantum Hall transport \cite{appugliese2022breakdown, Enkner_2025}, 
and {to enable} cavity control via ultrastrong coupling in van der Waals (vdW) heterostructures~\cite{kipp2025cavity}. On the theory side, model Hamiltonian studies predict cavity-mediated long-range electron pairing \cite{Jacksh_photoinduced}, resonantly enhanced photo-induced superconductivity \cite{eckhardt2024resonantly}, cavity light-matter entanglement \cite{passetti2023cavity}, and photonic probes of electronic phases \cite{nambiar2025diagnosing}, among others~\cite{sentef2018polaritonic,liu2025cavitymediated}. Specifically for dispersion interactions, ab initio studies of molecular systems have shown that cavities can modify these interactions in  {vdW-bound systems}~\cite{cavityvdwhaugland2025,tasci2025}.
%and potential energy surfaces \cite{weber2026magnetic}.

Connecting these observations and predictions to specific materials in experimentally realizable cavity setups requires ab initio methods that treat electrons and photons on the same footing \cite{polaritonicchemreview2023, polaritonicchemreview_ruggenthaler2023}. Quantum electrodynamical density-functional theory (QEDFT)~\cite{tokatly2013,ruggenthaler2014} provides such a framework, in which the effects of the cavity are captured by the choice of the electron-photon exchange-correlation functional. For extended systems, the photon-free electron-photon functional{~\cite{lu2024cavitymgb2}} captures cavity-induced short-range modifications of the electron density and band structure \cite{Liu_QEDFTvdW, I-Te_epLDA}, and recent formal work has established the validity of the long-wavelength, few-mode description under periodic boundary conditions \cite{cavitymaterialsSvendsen2025}. %polaritonicblochstheoremlongwavelengthbruno2026
The photon many-body dispersion (pMBD) functional~\cite{tasci2025, tasci2025super}, in turn, provides a non-perturbative description of cavity-modified long-range dispersion interactions, but has so far been limited to finite systems. Hence, no ab initio theory of cavity-modified dispersion interactions in extended materials exists to date: for periodic systems, dispersion is included only through pairwise-additive corrections that remain unaffected by the cavity \cite{Liu_QEDFTvdW}, leaving a gap precisely for the non-additive, long-range correlations that can dominate the interlayer binding of layered vdW materials~{\cite{Ambrosetti2016, tkatchenko_graphene2015}} and
that the cavity modifies most strongly~\cite{cavityvdwhaugland2025}.

In this Letter, we close this gap by introducing a periodic formulation of the pMBD functional. Within the long-wavelength approximation, we predict cavity-modified stacking for bilayer hBN and graphene, as well as coupling-dependent equilibrium interlayer distances and layer breathing mode (LBM) frequencies. {Our results show that cavity-mediated long-range interactions can serve as a tuning knob for precisely those structural properties (stacking order and interlayer distance) that are relevant to emergent phases of moir\'e and twisted heterostructures, from stacking ferroelectricity in bilayer hBN~\cite{yasuda2021stacking} to superconductivity in magic-angle graphene~\cite{cao2018unconventional}.}

\textit{Theory:} The pMBD Hamiltonian in the momentum gauge and long-wavelength approximation is given as follows {\cite{tasci2025super, I-Te_epLDA} (see the Supplemental {Material} for the derivation)}
{%
\begin{align}
       \hat{H}&=\frac{1}{2}\sum_{i=1}^{N_a}\sum_{a=1}^{3}\left[\left(\hat\pi_{ia}-\frac{\omega_{ia}\sqrt{\alpha_{ia}}}{c}\hat{A}_a\right)^2 + \omega^2_{ia}\hat\chi_{ia}^{2}\right]\nonumber\\
       &+\frac{1}{2}\sum_{i,j=1}^{N_a}\sum_{a,b=1}^3\omega_{ia}\omega_{jb}\sqrt{\alpha_{ia}\alpha_{jb}}\,\hat\chi_{ia}T_{\text{LR},ij}^{ab}\hat\chi_{jb}\nonumber\\
       &+ \frac{1}{2}\sum_{\alpha=1}^{N_p}\left(\hat{p}_{\alpha}^{2} +\omega_{\alpha}^2\hat{q}_{\alpha}^2\right) \label{eq:rpa-ham-mom}
\end{align}}

Here, each atom $i$ is associated with three (for each spatial direction $a$) quantum harmonic oscillators (QHO) with frequency $\omega_{ia}$, charges $e_{ia}$, polarizabilities $\alpha_{ia} = {e^2_{ia}}/(m_i \omega_{ia}^2)$ and (mass-weighted) atomic displacement (momentum) $\hat\chi_{ia}$ ($\hat\pi_{ia}$). The photonic subsystem is described by cavity frequencies $\omega_\alpha$ and the photon displacement (momentum) coordinate $\hat{q}_\alpha$ ($\hat{p}_\alpha$). Light-matter coupling is introduced via minimal coupling with the vector potential $\hat{\textbf A} = c\sum_{\alpha=1}^{N_p}\boldsymbol\lambda_{\alpha}  \hat{q}_\alpha$. The light-matter coupling strength, $\boldsymbol\lambda_{\alpha}${,} has a magnitude connected to the effective volume $V_\alpha$ by $\lambda_{\alpha} = \sqrt{\frac{4\pi}{V_\alpha}}$ \cite{climent2019plasmonic,I-Te_epLDA}\footnote{The coupling coefficient can be written either as $e_{ia}/(c\sqrt{m_i})$ or as $\omega_{ia}\sqrt{\alpha_{ia}}/c$, which are equal by the definition of $\alpha_{ia}$; the latter is preferable because it involves only quantities that actually enter the MBD model.}. Finally, the dipole-dipole interaction between the atoms is described by the dipole interaction tensor {$T_{\text{LR},ij}^{ab}$}~\cite{ambrosetti2014long}.

%After expanding the kinetic term, we can rewrite the Hamiltonian as:

%\begin{align}
%\label{eq:rpa-ham-mom_expand}
%       \hat{H}_\text{pMBD,M}&=\frac{1}{2}\sum_{i=1}^{N_a}\sum_{a=1}^{3} \pi_{ia}^2 + \omega^2_{ia}\chi_{ia}^{2}\nonumber\\
%       &+\frac{1}{2}\sum_{i,j=1}^{N_a}\sum_{a,b=1}^3 e_{ia}e_{jb}\chi_{ia}T_{\text{LR},ij}^{ab}\chi_{jb}\nonumber\\
%       &-\frac{1}{c}\hat{\textbf{A}} \cdot \hat{\textbf{J}}\nonumber\\
%       &+ \frac{1}{2}\sum_{\alpha=1}^{N_p}\left(\hat{p}_{\alpha}^{2} +\omega_{\alpha}^2\hat{q}_{\alpha}^2\right) + \frac{Q}{2c^2}\hat{\textbf{A}}^2
%\end{align}

%where $Q=\sum_{i=1}^{N_a}\sum_{a=1}^{3}e_{ia}^2$. It becomes clear that the vector potential couples to the paramagnetic current operator $\hat{\textbf{J}}=\sum_{i=1}^{N_a}\sum_{a=1}^{3} e_{ia} \pi_{ia} \boldsymbol{u}_a$, where $\boldsymbol{u}_a$ is the cartesian unit vector pointing in direction $a$.

To adapt this general Hamiltonian into a more convenient formulation for periodic systems, we  introduce the mapping onto a periodic lattice~\cite{libmbd,bucko2016many}. To do so, we assume a periodic solid %with Born--von Karman (BvK) boundary conditions 
on a supercell consisting of $N_c$ periodic cells, and we describe $\textbf R$ as the Bravais lattice vector of cell {$\boldsymbol n$} and the coordinate $\textbf R_i$ describes the position of atom $i$ in the unit cell. The atomic displacement (momentum) coordinate then becomes $\chi_{ia,\textbf R}$ ($\pi_{ia,\textbf R}$). The Hamiltonian then becomes

{%
\begin{align}
\label{eq:ham_periodic}
       \hat{H}&=\frac{1}{2}\sum_{\textbf R}\sum_{ia}\left[\left(\hat\pi_{ia,\textbf R}-\frac{\omega_{ia}\sqrt{\alpha_{ia}}}{c}\hat{A}_a\right)^2+\omega^2_{ia}\hat\chi_{ia,\textbf R}^{2}\right]\nonumber\\
       &+\frac{1}{2}\sum_{\textbf{R},\textbf{R}'}\sum_{ij}\sum_{ab}\omega_{ia}\omega_{jb}\sqrt{\alpha_{ia}\alpha_{jb}}\,\hat\chi_{ia,\textbf R}\nonumber \\
       &\times T_{\text{LR},ij}^{ab}(\textbf R' + \textbf R_j - \textbf R -\textbf R_i)\,\hat\chi_{jb,\textbf R'}\nonumber\\
       &+ \frac{1}{2}\sum_{\alpha=1}^{N_p}\left(\hat{p}_{\alpha}^{2} +\omega_{\alpha}^2\hat{q}_{\alpha}^2\right)
\end{align}}

%where $\hat{\textbf{J}}_\textbf R = \sum_{\textbf R}\sum_{ia} e_{ia} \pi_{ia, \textbf R} \boldsymbol{u}_a$ and the diamagnetic term acquires a factor of $N_c$ to account for all of the charges across the periodic cells. 
%The real-space Hamiltonian can be transformed into a momentum space Hamiltonian by applying the Fourier Transform on the matter operators, for which deta

Now, we can transform the atomic operators into reciprocal space ($\sum_\textbf{R} \rightarrow \sum_\textbf{q}$) by using  $\chi_{ia,\textbf R} = \frac{1}{\sqrt N_c}\sum_\textbf q e^{i\textbf q\left(\textbf R + \textbf R_i\right)} \chi_{ia}(\textbf q)
 $ as well as the identity
 $\sum_\textbf R e^{i \left(\textbf q - \textbf{q}'\right)\cdot \textbf R} = N_c \delta_{\textbf q \textbf q'}$. Therefore, for the full Hamiltonian in $\textbf{q}$-space we get

{%
\begin{align}
\label{eq:ham_gamma}
   \hat{H} &= \hat{H}_\text{MBD} +\frac{1}{2}\sum_{\alpha=1}^{N_p}\left(\hat{p}_{\alpha}^{2} +\omega_{\alpha}^2\hat{q}_{\alpha}^2\right)
    \nonumber\\
   &- \frac{\sqrt{N_c}}{c}\hat{\textbf{A}} \cdot \hat{\textbf{J}}_{\textbf{q}=0} + \frac{N_c}{2 c^2}\sum_{a}\Big(\sum_{i}\omega_{ia}^2\alpha_{ia}\Big)\hat{A}_a^2.
\end{align}}

{Here, we introduced the} {paramagnetic current operator $\hat{J}_{\textbf{q}=0, a} = \sum_{i=1}^{N_a}\omega_{ia}\sqrt{\alpha_{ia}}\,\hat\pi_{ia}\left(\textbf{q}=0\right)$}, which, in the long-wavelength approximation, couples to the photon mode only at the $\Gamma$-point ($\textbf q = 0$) (in-phase collective dipole oscillation). In the above equation, $\hat{H}_\text{MBD}$ is the regular MBD Hamiltonian~\cite{libmbd} in $\textbf{q}$-space given by

{%
\begin{align}
\label{eq:ham_mbd_qspace}
\hat{H}_\text{MBD} &= \frac{1}{2} \sum_\textbf{q} \left[\sum_{ia}\Big(\hat\pi_{ia}(\textbf{q})\hat\pi_{ia}(-\textbf{q})+\omega^2_{ia}\hat\chi_{ia}(\textbf{q})\hat\chi_{ia}(-\textbf{q})\Big)\right.\nonumber\\
&\left.+\sum_{ij}\sum_{ab}\omega_{ia}\omega_{jb}\sqrt{\alpha_{ia}\alpha_{jb}}\,\hat\chi_{ia}(\textbf q)\,T_{\text{LR},ij}^{ab}(\textbf{q})\,\hat\chi_{jb}(-\textbf{q}) \right]
\end{align}}

In Eq.~{(}\ref{eq:ham_mbd_qspace}{)}, the dipole-dipole interaction tensor is expressed as {$T_{\text{LR},ij}^{ab}(\textbf{q}) = \sum_{\boldsymbol n}^{'} T_\text{LR}^{ab}(\textbf{R}_{\boldsymbol{n}ij})e^{-i\textbf{q} \cdot \textbf{R}_{\boldsymbol{n}ij}}$ with $\textbf{R}_{\boldsymbol{n}ij} = \textbf{R}_j + \textbf{R}_{\boldsymbol n} - \textbf{R}_i$, where $\textbf{R}_{\boldsymbol n}$ is the lattice vector of periodic cell $\boldsymbol n$~\cite{libmbd}}, and $\sum^{'}$ indicates that the sum is over an infinite number of periodic cells, with exclusion of the $\boldsymbol n = 0$ and $i = j$ terms. As a result, in the unit cell, and due to the terms $\sqrt{N_c} \hat{\textbf A}$ and $N_c \hat{\textbf A}^2$ in Eq.~{(}\ref{eq:ham_gamma}{)}, the light-matter coupling is enhanced by $\sqrt{N_c}$, {where $N_c$ is} the number of unit cells that coherently couple to the cavity mode~\cite{fan2026unified}. We thus can define an effective coupling strength as $\boldsymbol\lambda_{\alpha,\text{eff}} = \sqrt{N_c}\boldsymbol\lambda_\alpha$.

The total cavity-modified (exchange-correlation) energy is then obtained from diagonalizing this Hamiltonian at each \textbf{q} point:

\begin{align}
\label{eq:finite_energy}
       E_\text{xc}(\textbf{q})&=\frac{1}{2}\sum_{k=1}^{3N_a + N_p} \Omega_k(\textbf{q}) - \frac{1}{2} \sum_{i=1}^{N_a}\sum_{a=1}^{3}\omega_{ia} -\frac{1}{2}\sum_{\alpha=1}^{N_p} \omega_\alpha
\end{align}

where $\Omega_k^2(\textbf{q})$ are the eigenvalues of the dynamical matrix associated with the Hamiltonian in Eq.~(\ref{eq:ham_gamma}). The energy for the total periodic Hamiltonian can now be obtained as an integral of Eq.~(\ref{eq:finite_energy}) over the first Brillouin zone (FBZ):

\begin{align}
\label{eq:periodic_energy}
       E_\text{xc}&=\frac{1}{\Omega_\text{FBZ}}\int_{\text{FBZ}} \text{d}\textbf{q} ~ E_\text{xc}\left(\textbf{q}\right) \rightarrow \frac{1}{N_q}\sum_\textbf{q} E_\text{xc}(\textbf{q})
\end{align}

where $\Omega_\text{FBZ}$ is the FBZ volume and $N_q$ is the number of \textbf{q}-points used to discretize the FBZ for numerical calculation of the integral. We evaluate this integral as a finite sum over a chosen \textbf{q}-point grid, where the light-matter coupling then only acts on the $\Gamma$-point. {We define the cavity-mediated energy $\Delta E_\text{xc}$ at a specific coupling strength as the photonic contribution to $E_\text{xc}$, which can be calculated as the energy difference in the finite coupling strength simulation and the  $\lambda_\alpha = 0$ a.u. simulation. Since the cavity only influences $E_\text{xc}(\textbf{q})$ in Eq.~{(}\ref{eq:finite_energy}{)} at the $\Gamma$ point, we find $\Delta E_\text{xc} = \frac{1}{N_q} \left(E_\text{xc}(\textbf{q}=0) - E_\text{xc}(\textbf{q}=0, \boldsymbol\lambda_\alpha = 0) \right)$, i.e.{,} we only have to subtract the respective energy values at $\textbf{q}=0$. Thus, the additional computational cost of including electron-photon interactions is essentially independent of $N_c$: compared to outside the cavity, the cavity-modified energy only requires diagonalizing one additional  $(3N_a+N_p)$-dimensional dynamical matrix at the $\Gamma$ point.} In the {Supplemental Material}, we show how the Power-Zienau-Woolley (PZW) transformation can be used to transform the momentum gauge Hamiltonian into the length gauge. The length gauge Hamiltonian {is} used in the following for all numerical simulations.

%\section{Results}

%\subsection{Binding Curves for Bilayer Graphene and hBN}

\begin{figure}[th]
    \centering
    \includegraphics[scale=0.34]{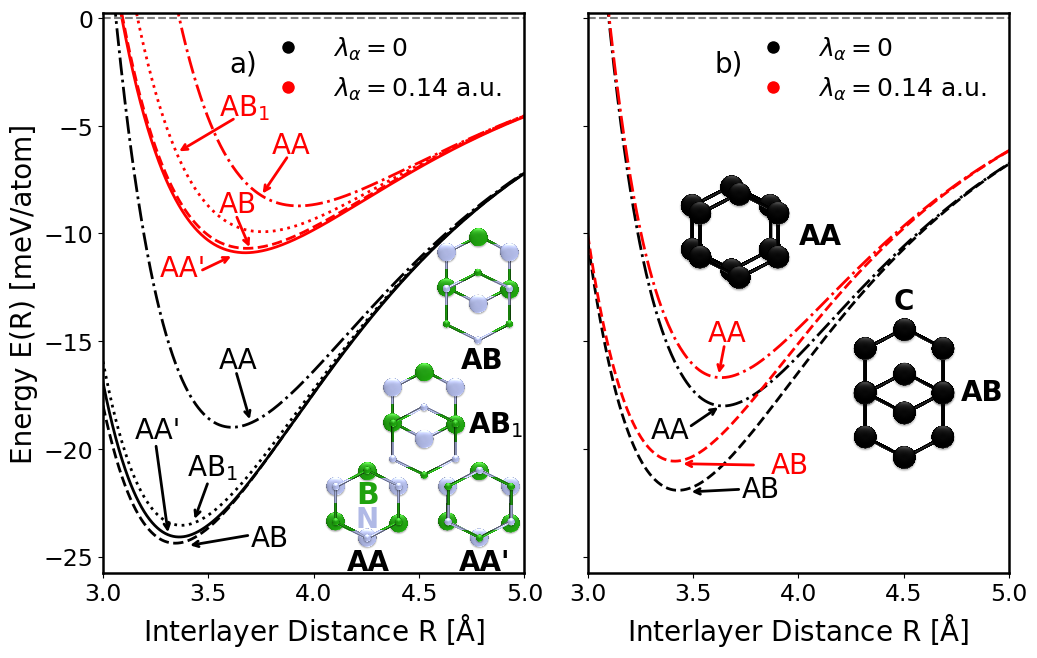}
    \caption{Binding curves for bilayer hBN and bilayer graphene. The y-axis is the sum of the DFT energy and $E_\text{xc}$ with respect to an offset at an interlayer distance of 10 $\text{\AA}$. a) Binding curves for bilayer hBN in AA, AB, {AB$_1$,} and AA' stacking. The black curves are the energies outside the cavity ($\lambda_\alpha = 0$~a.u.). The red curves are the energies in the cavity with $\omega_\alpha = 2$ eV and $\lambda_\alpha=0.14$ a.u., with a single cavity mode polarized in the out-of-plane direction.  
    b) Binding curves for bilayer graphene in AA and AB stacking.}
    \label{fig:disscurves} 
\end{figure}

\textit{Results:} We now implement this framework to study the binding curves of 2D vdW materials. For all of the following calculations, we calculate the DFT energy and Hirshfeld coefficients with the VASP software package \cite{vasp_PW, vasp_PAW} using the PBE exchange-correlation functional~\cite{PBE} and a \textbf{k}-point grid of $31 \times 31 \times 1$, where the \textbf{k}-point grid refers to the discretization of the FBZ for the electronic wavefunctions of the DFT calculation. We use a plane-wave cutoff of 520 eV with an out-of-plane  cell height of 20 $\text{\AA}$. In the {Supplemental Material}, we validate our framework by showing the equivalence in cavity mediated photon energies between calculations done on a unit cell and those done on a supercell. In particular, we show that a supercell calculation sampled at the $\Gamma$-point using $N_c$ cells with a coupling strength of $\lambda_{\alpha,\text{eff}}=\lambda_\alpha$ is equivalent to a unit cell calculation with a coupling strength of $\lambda_{\alpha,\text{eff}}=\sqrt{N_c}\lambda_\alpha$ sampled over a \textbf{q}-point grid consisting of $N_c$ points (provided that the number of cells/\textbf{q}-points  {is} the same in each direction{)}. In this case, we recover the equivalence between the size of the supercell and the number of \textbf{q}-points. Thus, a crystal in which $N_c$ cells couple to a cavity mode at the $\Gamma$-point is efficiently calculated using a unit cell sampled over $N_c$ \textbf{q}-points.  For our calculations, we choose $N_c=15$ unit cells to couple to a single, out-of-plane cavity mode by using a $5\times3\times1$ \textbf{q}-point grid sampled over a hexagonal unit cell, wherein the cavity mode has a frequency of $\omega=2$ eV. This setup  {roughly} corresponds to a small sample size of $1.3$ nm $\times$  $0.75$ nm. For this $5\times3\times1$ \textbf q-point grid, we then obtain the pMBD energy using Eq.~(\ref{eq:periodic_energy}) and add the converged MBD energy obtained on a \textbf{q}-point grid
of $37\times37\times1$. The converged $E_\text{xc}$ is then added to the DFT energy to generate the binding curves~\cite{libmbd,tasci2025}. {In the following, we show results for $\lambda_\alpha$ between 0 and 0.2~a.u.: while the upper end of this range is chosen high to exemplify the arising features, values up to $\lambda_\alpha=0.1$~a.u., corresponding to an effective mode volume of $V_\alpha\approx0.19$~nm$^3$~\cite{pavosevic2022cavity}, are within current reach of picocavities~\cite{benz2016single,richardlacroix2020direct,liu2021atomic,griffiths2021locating,urbieta2018atomic,wu2021bright, Baumberg2022picocavities}.} In the {Supplemental Material}, we provide more details on the interpretation of the coupling strength.

Applying our method to bilayer hBN, we calculate the binding curves in the AA, AB, AB$_1$, and AA' stacking configurations, which we plot in Fig.~\ref{fig:disscurves}a. For interlayer distances ranging from 3.0 $\text{\AA}$ to 5.1 $\text{\AA}$, we plot the energy with respect to an offset, which we set to be the energy of the respective bilayer at an interlayer distance of $10~\text{\AA}$. We find that outside of the cavity (black curves), AB is the most stable stacking configuration. With our implementation, we obtain outside a cavity ($\lambda_\alpha = 0$ a.u.) equilibrium binding energies of 24.1 meV/atom and 24.4 meV/atom at equilibrium interlayer distances of {3.36} $\text{\AA}$ and {3.34} $\text{\AA}$ for AA' and AB stacking, respectively, thus predicting AB as the preferred stacking order. {Experimental results  {indicate} AA' to be the preferred stacking configuration~\cite{stackpref1, stackpref2, stackpref3}, as well as calculations using the vdW-DF2 functional~\cite{constantinescu2013stacking} and MP2~\cite{lebedev2016interlayer}; however,} as the predicted energy differences between stackings are smaller than
0.5~meV{/atom}, these numbers are pushing the accuracy limits of DFT. {We note that our calculations for AA' and AB$_1$ stacking are in line with previous PBE+MBD simulations~\cite{tkatchenko_graphene2015}, both in the prediction of binding energies and equilibrium interlayer distances (within $0.7$~meV/atom and $0.05~\text{\AA}$, respectively) as well as in preferred stacking arrangements}. 

When the cavity is turned on and the coupling strength is increased, the preferred stacking is eventually flipped, where AA' becomes the most stable stacking configuration. Here, we plot the binding curves for a coupling strength of $\lambda_\alpha = 0.14$ a.u. (red curves), where we show clearly that the minimum energy for AA' stacking becomes lower than that of AB. We note that, while the underlying stacking energy differences are very close, 
{our main conclusion concerns} the differential cavity response of the stacking configurations, which grows with their polarizability (see Supplemental Material). In Fig.~\ref{fig:disscurves}b, we plot the binding curves for bilayer graphene in the AA and AB stacking configurations. The equilibrium binding energies and interlayer distances outside of the cavity (black curves) are 18.0 meV/atom and 3.64 $\text{\AA}$ in AA stacking, and 21.9 meV/atom and 3.43 $\text{\AA}$ in AB stacking. These reported binding energies and equilibrium interlayer distances for bilayer graphene at $\lambda_\alpha = 0$ a.u. are in {reasonable} agreement with previous results for PBE+MBD \cite{tkatchenko_graphene2015}. Previously, it has been shown that the magnitude of the cavity modification to the exchange-correlation dispersion energy depends on the polarizability of the material~\cite{flick2022simple}. Thus, the cavity induces a greater shift in the energy for stacking configurations {that} are predicted by MBD to be more {polarizable}. For this reason, since bilayer graphene is less polarizable than bilayer hBN, we see less of a change in the energies with respect to coupling strength{,} and we do not observe a change in preferred stacking behavior. It follows that as the polarizability of a bilayer increases, the rate of change of its energy with respect to coupling strength also increases. This is most obvious in the case of bilayer hBN, where the change in energy for AA' and AB stacking is far greater than that of AA. In the {Supplemental Material}, for different stacking configurations of bilayer graphene and bilayer hBN, we plot the energy with respect to coupling strength subtracted by the energy outside of the cavity and we show that the photonic contribution to the energy increases faster for more polarizable bilayers. Thus, for any 2D vdW material, as the polarizability of the stacking configurations increases, the coupling strength at which preferential stacking behavior is modified decreases.

\begin{figure}[h]
    \centering
    \includegraphics[scale=0.35]{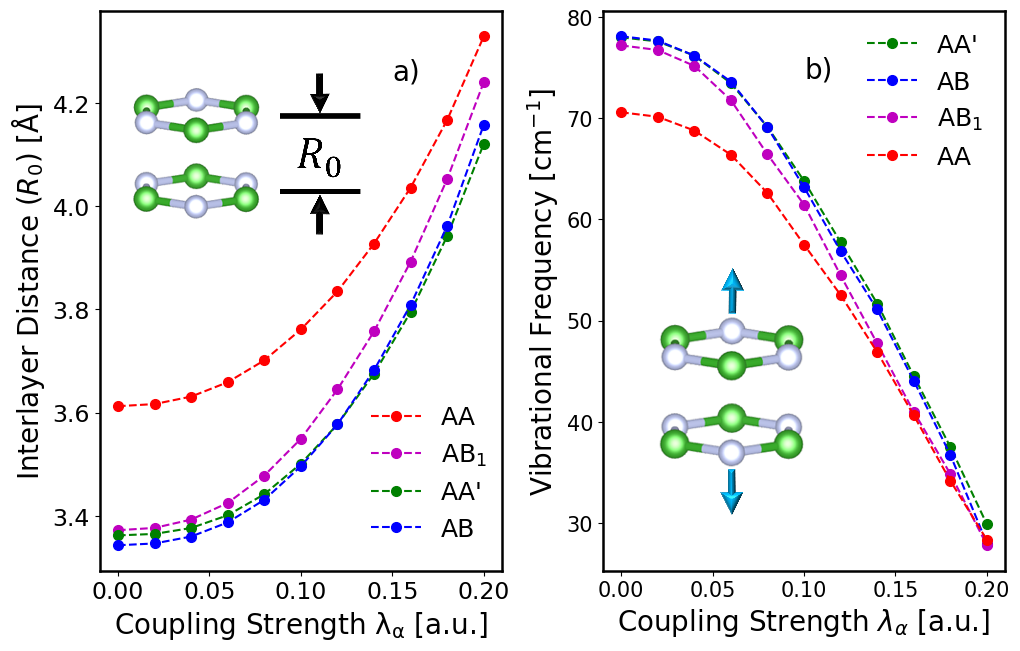}
    \caption{Interlayer distances (a) and layer breathing mode frequencies (b) as a function of coupling strength for bilayer hBN in a cavity with a single cavity mode polarized in the out-of-plane direction with a cavity frequency of $\omega = 2$ eV.}
    \label{fig:freq}
\end{figure}

In Fig.~\ref{fig:disscurves}, we find that increasing the coupling strength leads to decreased binding energies, which indicates that the interaction of the dipole fluctuations with the cavity mode has the effect of decreasing interlayer attraction. This decreased attraction implies that the equilibrium interlayer distances {increase, while the} LBM frequencies  decrease. We calculate the LBM frequencies, interlayer distances, and minimum energies at each coupling strength by fitting the binding energy curves from Fig.~\ref{fig:disscurves} to a third{-}order anharmonic potential. In Fig.~\ref{fig:freq}a, we plot the equilibrium interlayer distance with respect to coupling strength in the AA, AB, AB$_1$, and AA' stacking configurations. We show that increasing coupling strength leads to a significant increase in the equilibrium interlayer distance. {As shown in Fig.~\ref{fig:freq}a, the equilibrium interlayer distance increases by $0.14$--$0.18~\text{\AA}$ at $\lambda_\alpha=0.1$ a.u. and by up to $0.9~\text{\AA}$ at $\lambda_\alpha=0.2$ a.u., with all four stacking configurations expanding at a comparable rate. This expansion directly reflects the cavity-induced weakening of the interlayer attraction: as the binding wells in Fig.~\ref{fig:disscurves}a become shallower with increasing coupling strength, their minima shift to larger interlayer distances.}

This is in contrast to short-range effects obtained through photon-free electron-photon functionals, which predict decreased equilibrium interlayer distances in 2D vdW materials due to cavity modified electron densities when coupling to a single out-of-plane cavity mode~\cite{Liu_QEDFTvdW}. In Fig.~\ref{fig:freq}b, we plot the LBM frequencies with respect to coupling strength for each stacking configuration. For AA' and AB stacking, the LBM at $\lambda_\alpha=0$ a.u. (78.0 cm$^{-1}$ and 78.1 cm$^{-1}$, respectively) is close to what is reported from DFT calculations~\cite{rani2025nonlinear}. We show that the frequencies decrease significantly with increasing coupling strength, decreasing by $18\%$ at $\lambda_\alpha=0.1$ a.u. A similar effect is shown for bilayer graphene in the {Supplemental Material}. Additionally, the LBM frequency for AB$_1$ stacking drops off at a quicker rate such that it has the lowest frequency at $\lambda_\alpha=0.2$ a.u. despite having nearly the same frequency as AA' and AB stacking at $\lambda_\alpha=0$ a.u.

\textit{Summary and Conclusions:} {In this Letter, we introduce a periodic formulation of the pMBD exchange-correlation functional within QEDFT, providing ab initio access to cavity-modified dispersion interactions in extended materials. In the long-wavelength approximation, the cavity mode couples exclusively to the $\Gamma$-point, the in-phase collective dipole oscillation of all unit cells, and the equivalence between unit cell and supercell calculations yields an efficient $\textbf{q}$-point sampling scheme (see Supplemental Material). For bilayer hBN and bilayer graphene, we find that coupling to an out-of-plane cavity mode decreases the interlayer attraction: equilibrium interlayer distances increase by approximately $0.15~\text{\AA}$ and LBM frequencies soften by $18$--$20\%$ already at $\lambda_\alpha=0.1$ a.u., and the near-degeneracy of AB and AA' stacking in bilayer hBN is lifted with increasing coupling. Since these modifications grow with the polarizability, vdW heterostructures combining layers of dissimilar polarizability could exhibit stacking control also at weaker coupling strengths.} Looking forward, extending the method to {include} finite-momentum ($\textbf q \neq 0$) coupling with momentum-selective control of dispersion interactions and multimode cavities is particularly appealing for hBN, which is {a} hyperbolic material with phonon-polariton modes and deeply sub-wavelength field confinement~\cite{caldwell2019photonics}. Finally, applying the periodic pMBD framework to moir\'e and twisted bilayers opens the door to cavity control of stacking-dependent phases in twistronics{~\cite{carr2017twistronics,andrei2020graphene}}.

\textit{Data availability:} All results have been generated using the libphotonmbd software package~\cite{libphotonmbd}. Data and plot scripts are available on Github~\cite{datasets_used_in_manuscripts}.

\textit{Acknowledgments:} We acknowledge funding from the Defense Advanced Research Projects Agency (DARPA), and startup funding from the City College of New York. All calculations were performed using the computational facilities of the Flatiron Institute. The Flatiron Institute is a division of the Simons Foundation.

\bibliography{refs} % Produces the bibliography via BibTex.

\end{document}

\title{Supplemental Material: Modifying van der Waals Materials via Cavity Vacuum Fluctuations}

\author{Mohammad Hassan}
\affiliation{Department of Physics, City College of New York, New York, NY 10031, USA}
\affiliation{Department of Physics, The Graduate Center, City University of New York, New York, NY 10016, USA}
\author{Cankut Tasci}
\affiliation{Department of Physics, City College of New York, New York, NY 10031, USA}
\affiliation{Department of Physics, The Graduate Center, City University of New York, New York, NY 10016, USA}
\author{Leon Orlov-Sullivan}
\affiliation{Department of Physics, City College of New York, New York, NY 10031, USA}
\author{Leonardo A. Cunha}
\affiliation{Center for Computational Quantum Physics, Flatiron Institute, New York, 10010  NY,  USA}
\affiliation{Department of Chemistry and Biochemistry, Bates College, Lewiston, ME 04240, USA}
\author{Johannes Flick}
  \email[Electronic address:\;]{jflick@ccny.cuny.edu}
\affiliation{Department of Physics, City College of New York, New York, NY 10031, USA}
\affiliation{Department of Physics, The Graduate Center, City University of New York, New York, NY 10016, USA}
\affiliation{Center for Computational Quantum Physics, Flatiron Institute, New York, 10010  NY,  USA}

\date{\today}

\maketitle

\section{Derivation of the Periodic photon MBD (pMBD) Hamiltonian in Momentum and Length Gauges}

\subsection{General Setup}

{In the many-body dispersion (MBD) model~\cite{tkatchenko2012accurate}, each atom $i = 1,\dots,N_a$ is represented by three charged quantum harmonic oscillators (QHO), one for each Cartesian direction $a = 1,2,3$, with frequency $\omega_{ia}$, charge $e_{ia}$, mass $m_i$, and static polarizability
\begin{align}
\label{eq:alpha_def}
\alpha_{ia} = \frac{e_{ia}^2}{m_i\,\omega_{ia}^2}.
\end{align}
We denote the Cartesian displacement of oscillator $(i,a)$ away from the atomic position $\textbf{R}_i$ by $\xi_{ia}$ and its conjugate momentum by ${p}_{ia}$. The oscillators interact through the long-range dipole-dipole tensor $T^{ab}_{\text{LR},ij}$, and couple to $N_p$ cavity modes through minimal coupling. In the long-wavelength approximation, the vector potential is spatially uniform,
\begin{align}
\label{eq:A_def}
\hat{\textbf{A}} = c\sum_{\alpha=1}^{N_p}\boldsymbol{\lambda}_{\alpha}\,\hat{q}_{\alpha},
\qquad
\lambda_{\alpha} = \sqrt{\frac{4\pi}{V_\alpha}},
\end{align}
where the light-matter coupling strength $\boldsymbol{\lambda}_\alpha$ points along the cavity polarization direction and is connected to the effective mode volume $V_\alpha$, and the photon displacement (momentum) operators $\hat{q}_\alpha$ ($\hat{p}_\alpha$) obey $[\hat{q}_\alpha,\hat{p}_{\alpha'}] = i\delta_{\alpha\alpha'}$. The minimal-coupling Hamiltonian then reads~\cite{tasci2025super}
\begin{align}
\label{eq:min_coupling}
\hat{H} &= \sum_{i=1}^{N_a}\sum_{a=1}^{3}\left[\frac{1}{2m_i}\left(\hat{p}_{ia}-\frac{e_{ia}}{c}\hat{A}_a\right)^2 + \frac{1}{2}m_i\omega_{ia}^2\hat\xi_{ia}^2\right]\nonumber\\
&+ \frac{1}{2}\sum_{i,j=1}^{N_a}\sum_{a,b=1}^{3}e_{ia}e_{jb}\,\hat\xi_{ia}T^{ab}_{\text{LR},ij}\hat\xi_{jb}\nonumber\\
&+ \frac{1}{2}\sum_{\alpha=1}^{N_p}\left(\hat{p}_{\alpha}^2+\omega_\alpha^2\hat{q}_{\alpha}^2\right).
\end{align}
Introducing mass-weighted coordinates and momenta, $\hat\chi_{ia} = \sqrt{m_i}\,\hat\xi_{ia}$ and $\hat\pi_{ia} = \hat{p}_{ia}/\sqrt{m_i}$, with $[\hat\chi_{ia},\hat\pi_{jb}] = i\delta_{ij}\delta_{ab}$, the charges only appear in the mass-weighted combination $e_{ia}/\sqrt{m_i} = \omega_{ia}\sqrt{\alpha_{ia}}$, which follows from Eq.~(\ref{eq:alpha_def}). Therefore, Eq.~(\ref{eq:min_coupling}) becomes the momentum gauge pMBD Hamiltonian
\begin{align}
\label{eq:rpa-ham-mom}
       \hat{H}&=\frac{1}{2}\sum_{i=1}^{N_a}\sum_{a=1}^{3}\left[\left(\hat\pi_{ia}-\frac{\omega_{ia}\sqrt{\alpha_{ia}}}{c}\hat{A}_a\right)^2 + \omega^2_{ia}\hat\chi_{ia}^{2}\right]\nonumber\\
       &+\frac{1}{2}\sum_{i,j=1}^{N_a}\sum_{a,b=1}^3\omega_{ia}\omega_{jb}\sqrt{\alpha_{ia}\alpha_{jb}}\,\hat\chi_{ia}T_{\text{LR},ij}^{ab}\hat\chi_{jb}\nonumber\\
       &+ \frac{1}{2}\sum_{\alpha=1}^{N_p}\left(\hat{p}_{\alpha}^{2} +\omega_{\alpha}^2\hat{q}_{\alpha}^2\right),
\end{align}
which corresponds to Eq.~(1) of the main text.}

\subsection{Periodic systems in the Momentum Gauge}

The derivation of the pMBD Hamiltonian in the momentum gauge for periodic systems begins with expanding the kinetic term in {Eq.~(\ref{eq:rpa-ham-mom})}:

\begin{align}
\label{eq:rpa-ham-mom_expand}
       \hat{H}&=\frac{1}{2}\sum_{i=1}^{N_a}\sum_{a=1}^{3} \left(\hat\pi_{ia}^2 + \omega^2_{ia}\hat\chi_{ia}^{2}\right)\nonumber\\
       &+\frac{1}{2}\sum_{i,j=1}^{N_a}\sum_{a,b=1}^3 \omega_{ia}\omega_{jb}\sqrt{\alpha_{ia}\alpha_{jb}}\,\hat\chi_{ia}T_{\text{LR},ij}^{ab}\hat\chi_{jb}\nonumber\\
       &-\frac{1}{c}\hat{\textbf{A}} \cdot \hat{\textbf{J}}\nonumber\\
       &+ \frac{1}{2}\sum_{\alpha=1}^{N_p}\left(\hat{p}_{\alpha}^{2} +\omega_{\alpha}^2\hat{q}_{\alpha}^2\right) + \frac{1}{2c^2}\sum_{i=1}^{N_a}\sum_{a=1}^{3}\omega_{ia}^2\alpha_{ia}\hat{A}_a^2
\end{align}

where all coefficients involve only quantities that enter the MBD model. It becomes clear that the vector potential couples to the {(mass-weighted)} paramagnetic current operator, which is defined as $\hat{{J_a}}=\sum_{i=1}^{N_a}\omega_{ia}\sqrt{\alpha_{ia}}\, \hat\pi_{ia}$.

To adapt this Hamiltonian for finite systems into a Hamiltonian for periodic systems, we need to introduce the mapping onto a periodic lattice. To do so, we will assume a system consisting of $N_c$ periodic cells, and describe $\textbf R$ as the Bravais lattice vector of cell $\boldsymbol n$ and the coordinate $\textbf R_i$ describes the position of atom $i$ in the unit cell. The atomic momentum (displacement) coordinate then becomes $\pi_{ia,\textbf R}$ ($\chi_{ia,\textbf R}$). Doing so the Hamiltonian then becomes

\begin{align}
\label{eq:ham_periodic_mom}
       \hat{H}&=\frac{1}{2}\sum_{\textbf R}\sum_{ia}\left(\hat\pi^2_{ia,\textbf R}+\omega^2_{ia}\hat\chi_{ia,\textbf R}^{2}\right)\nonumber\\
       &+\frac{1}{2}\sum_{\textbf{R},\textbf{R}'}\sum_{ij}\sum_{ab}\omega_{ia}\omega_{jb}\sqrt{\alpha_{ia}\alpha_{jb}}\,\hat\chi_{ia,\textbf R}T_{\text{LR},ij}^{ab}\nonumber \\
       &\times (\textbf R' + \textbf R_j - \textbf R -\textbf R_i)\,\hat\chi_{jb,\textbf R'}\nonumber\\
       &-\frac{1}{c}\hat{\textbf{A}} \cdot \hat{\textbf{J}}_\textbf R\nonumber\\
       &+ \frac{1}{2}\sum_{\alpha=1}^{N_p}\left(\hat{p}_{\alpha}^{2} +\omega_{\alpha}^2\hat{q}_{\alpha}^2\right) + \frac{N_c}{2c^2}\sum_{ia}\omega_{ia}^2\alpha_{ia}\hat{A}_a^2
\end{align}

where $\hat{{J}}_{a,\textbf R} = \sum_{\textbf R}\sum_{i} \omega_{ia}\sqrt{\alpha_{ia}}\, \hat\pi_{ia, \textbf R}$ and the diamagnetic term acquires a factor of $N_c$ to account for all of the charges across the periodic cells.
Now, we can transform the atomic operators into reciprocal space ($\sum_\textbf{R} \rightarrow \sum_\textbf{q}$) by using the Fourier transform for {the} position operator,
$\chi_{ia,\textbf R} = \frac{1}{\sqrt N_c}\sum_\textbf q e^{i\textbf q\left(\textbf R + \textbf R_i\right)} \chi_{ia}(\textbf q)
$, the momentum operator, $\pi_{ia,\textbf R} = \frac{1}{\sqrt N_c}\sum_\textbf q e^{-i\textbf q\left(\textbf R + \textbf R_i\right)} \pi_{ia}(\textbf q)
$ as well as the lattice orthogonality condition,
$\sum_\textbf R e^{i \left(\textbf q - \textbf{q}'\right)\cdot \textbf R} = N_c \delta_{\textbf q \textbf q'}$. {The opposite phase conventions for $\hat\chi$ and $\hat\pi$ guarantee hermiticity, $\hat\chi_{ia}(\textbf{q})^\dagger = \hat\chi_{ia}(-\textbf{q})$ and $\hat\pi_{ia}(\textbf{q})^\dagger = \hat\pi_{ia}(-\textbf{q})$, together with the canonical commutation relations $[\hat\chi_{ia}(\textbf{q}),\hat\pi_{jb}(\textbf{q}')] = i\delta_{ij}\delta_{ab}\delta_{\textbf{q}\textbf{q}'}$.}

{%
\begin{align}
\label{eq:ham_periodic_qspace}
       \hat{H}&=\frac{1}{2} \sum_\textbf{q} \left(\sum_{ia}\left(\hat\pi_{ia}(\textbf{q})\hat\pi_{ia}(-\textbf{q})\right. \right.\nonumber\\
       &\left.+\omega^2_{ia}\hat\chi_{ia}(\textbf{q})\hat\chi_{ia}(-\textbf{q})\right)\nonumber\\
       &\left.+\sum_{ij}\sum_{ab}\omega_{ia}\omega_{jb}\sqrt{\alpha_{ia}\alpha_{jb}}\,\hat\chi_{ia}(\textbf q)T_{\text{LR},ij}^{ab}(\textbf{q})\hat\chi_{jb}(-\textbf{q}) \right)\nonumber\\
       &-\frac{\sqrt{N_c}}{c}\hat{\textbf{A}} \cdot \hat{\textbf{J}}_{\textbf{q}=0}\nonumber\\
       &+ \frac{1}{2}\sum_{\alpha=1}^{N_p}\left(\hat{p}_{\alpha}^{2} +\omega_{\alpha}^2\hat{q}_{\alpha}^2\right) + \frac{N_c}{2c^2}\sum_{ia}\omega_{ia}^2\alpha_{ia}\hat{A}_a^2
\end{align}}

where $\hat{{J}}_{a,\textbf{q}=0} = \sum_{i} \omega_{ia}\sqrt{\alpha_{ia}}\, \hat\pi_{ia}\left(\textbf{q}=0 \right)$. {The factor $\sqrt{N_c}$ in the bilinear coupling arises from $\sum_{\textbf R}\hat\pi_{ia,\textbf R} = \sqrt{N_c}\,\hat\pi_{ia}(\textbf q = 0)$: only the $\textbf q = 0$ component of the current survives the lattice sum, so in the long-wavelength approximation the cavity mode couples exclusively to the in-phase ($\Gamma$-point) collective dipole oscillation. The lattice Fourier transform of the dipole tensor is given by $T^{ab}_{\text{LR},ij}(\textbf{q}) = {\sum_{\boldsymbol{n}}}' \,T^{ab}_{\text{LR}}(\textbf{R}_{\boldsymbol{n}ij})\,e^{-i\textbf{q}\cdot\textbf{R}_{\boldsymbol{n}ij}}$ with $\textbf{R}_{\boldsymbol{n}ij} = \textbf{R}_j + \textbf{R}_{\boldsymbol{n}} - \textbf{R}_i$, where $\textbf{R}_{\boldsymbol{n}} = \textbf{R}'-\textbf{R}$ is the relative lattice vector of periodic cell $\boldsymbol{n}$ and the primed sum excludes the $\boldsymbol{n}=0$,$i=j$ terms, consistent with Ref.~\cite{libmbd}.}

{The matter part of Eq.~(\ref{eq:ham_periodic_qspace}) is the regular MBD Hamiltonian in $\textbf{q}$-space~\cite{libmbd},
\begin{align}
\label{eq:ham_mbd_qspace}
\hat{H}_\text{MBD} &= \frac{1}{2} \sum_\textbf{q} \left(\sum_{ia}\left(\hat\pi_{ia}(\textbf{q})\hat\pi_{ia}(-\textbf{q})\right.\right.\nonumber\\
&\left.+\omega^2_{ia}\hat\chi_{ia}(\textbf{q})\hat\chi_{ia}(-\textbf{q})\right)\nonumber\\
&\left.+\sum_{ij}\sum_{ab}\omega_{ia}\omega_{jb}\sqrt{\alpha_{ia}\alpha_{jb}}\,\hat\chi_{ia}(\textbf q)T_{\text{LR},ij}^{ab}(\textbf{q})\hat\chi_{jb}(-\textbf{q}) \right),
\end{align}
such that the periodic momentum gauge Hamiltonian separates into the MBD part, the bare photon part, the bilinear light-matter coupling, and the diamagnetic term.}

\subsection{Power-Zienau-Woolley Transformation}

In quantum electrodynamics, the minimal coupling Hamiltonians in the momentum and length gauges are connected through the Power-Zienau-Woolley unitary transformation~\cite{faisal1987,craig1998,tokatly2013}, which is generally expressed as (in atomic units) $U_\text{PZW} = \exp(-{i}\hat{\textbf{D}} \cdot \hat{\textbf{A}}/c)$, where $\hat{\textbf{D}}$ is the total dipole moment operator. In order to apply the PZW transformation to the {momentum} gauge, the photonic quadratures must first be swapped by applying a quarter-period unitary evolution of the free photon mode. First, we define 
%$\hat{J}_a = \sum_{i=1}^{N_a} \omega_{ia}\sqrt{\alpha_{ia}}\hat{\pi}_{ia}(\textbf{q}=0)$ and 
$\hat{D}_a = \sum_{i=1}^{N_a} \omega_{ia}\sqrt{\alpha_{ia}}\hat{\chi}_{ia}(\textbf{q}=0)$, where $\hat{D}_a$ is the mass-weighted collective dipole operator of the unit cell. For clarity, we also consider a single cavity mode $\alpha$ with polarization vector $\boldsymbol{\lambda}_{\alpha} = (\lambda_{\alpha a})$, thus allowing us to express the vector potential as $\hat{A}_a = c\lambda_{\alpha a}\hat{q}_\alpha$. We can then rewrite the {momentum} gauge Hamiltonian {of Eq.~(\ref{eq:ham_periodic_qspace})} as

\begin{align}
\label{eq:ham_mom_compact}
\hat{H}&= \hat{H}_{\text{MBD}} + \frac{1}{2}\left(\hat{p}_{\alpha}^{2} +\omega_{\alpha}^2\hat{q}_{\alpha}^2 \right) \nonumber\\
       &- \sqrt{N_c}\hat{q}_\alpha \sum_a \lambda_{\alpha a} \hat{J}_a + \frac{N_c}{2}\sum_{ia} \lambda^2_{\alpha a} \omega_{ia}^{2} \alpha_{ia} \hat{q}^{2}_{\alpha}
\end{align}

Following Ref{s}.~\cite{tokatly2013, pellegrini2015optimized}, we interchange $p_\alpha$ and $q_\alpha$ next using the unitary transformation $\hat{W} = \exp(-i\frac{\pi}{2}\hat a^\dagger_\alpha \hat a_\alpha)$, where $a_\alpha$ and $a_\alpha^\dagger$ are the photon annihilation and creation operators, respectively. It follows {that}

\begin{align}
\hat{W}^\dagger\hat{q}_\alpha\hat{W} = \frac{\hat{p}_\alpha}{\omega_\alpha},
\qquad
\hat{W}^\dagger\hat{p}_\alpha\hat{W} = -\omega_\alpha\hat{q}_\alpha,
\end{align}

which leaves the bare photon Hamiltonian $\frac{1}{2}(\hat{p}_\alpha^2+\omega_\alpha^2\hat{q}_\alpha^2)$ invariant. Applying this transformation {to} Eq.~{(}\ref{eq:ham_mom_compact}{)} results in 

\begin{align}
\label{eq:step1}
\hat{W}^\dagger\hat{H}\hat{W}
&= \hat{H}'_{\text{MBD}} + \frac{1}{2}\hat{p}_\alpha^2 + \frac{1}{2}\omega_\alpha^2\hat{q}_\alpha^2 \nonumber\\
&+ \frac{1}{2}\sum_{ia}\left(\hat\pi_{ia}(\textbf{q}=0)-\frac{\sqrt{N_c}}{\omega_\alpha}\lambda_{\alpha a}\,\omega_{ia}\sqrt{\alpha_{ia}}\,\hat{p}_\alpha\right)^2
\end{align}
where $\hat{H}'_{\text{MBD}}$ collects all remaining terms of $\hat{H}_{\text{MBD}}$, i.e., the kinetic terms at $\textbf{q}\neq 0$ and all harmonic and dipole-dipole terms. %Expanding the square in Eq.~(\ref{eq:step1}) reproduces the cross term $-\frac{\sqrt{N_c}}{\omega_\alpha}\hat{p}_\alpha\sum_a\lambda_{\alpha a}\hat{J}_a$ and the transformed diamagnetic term $\frac{N_c}{2\omega_\alpha^2}\sum_{ia}\lambda_{\alpha a}^2\omega_{ia}^2\alpha_{ia}\,\hat{p}_\alpha^2$ exactly; the diamagnetic term is precisely the square-completion term. Eq.~(\ref{eq:step1}) has the structure of a minimal-coupling Hamiltonian with the roles of light and matter interchanged: the matter momenta at the $\Gamma$-point are shifted by an effective vector potential proportional to $\hat{p}_\alpha$. 

We next apply the PZW-type unitary
\begin{align}
\label{eq:pzw_unitary}
\hat{U}_{\text{PZW}} = \exp\left(i\,\frac{\sqrt{N_c}}{\omega_\alpha}\,\hat{p}_\alpha\sum_{a}\lambda_{\alpha a}\hat{D}_a\right),
\end{align}
the coupled-oscillator analogue of {the} PZW transformation~\cite{faisal1987,craig1998,tokatly2013}. Since the exponent is linear in $\hat\chi_{ia}(\textbf{q}=0)$ and $\hat{p}_\alpha$, the Baker-Campbell-Hausdorff expansion terminates after the first commutator and yields the exact relations
\begin{align}
\hat{U}_{\text{PZW}}^\dagger\,\hat\pi_{ia}(\textbf{q}=0)\,\hat{U}_{\text{PZW}} &= \hat\pi_{ia}(\textbf{q}=0) \nonumber\\
&+ \frac{\sqrt{N_c}}{\omega_\alpha}\lambda_{\alpha a}\,\omega_{ia}\sqrt{\alpha_{ia}}\,\hat{p}_\alpha,\\
\hat{U}_{\text{PZW}}^\dagger\,\hat{q}_\alpha\,\hat{U}_{\text{PZW}} &= \hat{q}_\alpha - \frac{\sqrt{N_c}}{\omega_\alpha}\sum_{a}\lambda_{\alpha a}\hat{D}_a,
\end{align}
while $\hat\chi_{ia}(\textbf{q})$ and $\hat{p}_\alpha$ remain unchanged.

\subsection{Length Gauge Hamiltonian}

Applied to Eq.~(\ref{eq:step1}), the shift of $\hat\pi_{ia}(\textbf{q}=0)$ removes the effective vector potential from the matter kinetic term, while the shift of $\hat{q}_\alpha$ transfers the light-matter coupling into the photonic potential:
\begin{align}
\label{eq:length_gauge}
\begin{aligned}
\hat{H}_{\text{L}} &= \hat{U}_{\text{PZW}}^\dagger\hat{W}^\dagger\,\hat{H}\,\hat{W}\hat{U}_{\text{PZW}}\\
&= \hat{H}_{\text{MBD}} + \frac{1}{2}\hat{p}_\alpha^2 \\
&+ \frac{1}{2}\left(\omega_\alpha\hat{q}_\alpha - \sqrt{N_c}\sum_{a}\lambda_{\alpha a}\sum_{i}\omega_{ia}\sqrt{\alpha_{ia}}\,\hat\chi_{ia}(\textbf{q}=0)\right)^2
\end{aligned}
\end{align}

which is the length gauge pMBD Hamiltonian for periodic systems, the direct generalization of the finite-system length gauge Hamiltonian of Ref.~\cite{tasci2025} with the replacement $\chi_{ia}\rightarrow\sqrt{N_c}\,\chi_{ia}(\textbf{q}=0)$. It becomes clear now that while the photon mode couples to the paramagnetic current operator in the {momentum} gauge, it couples to the total dipole moment operator in the length gauge. 

{For $N_p$ cavity modes, the transformation is applied to each mode separately, and the length gauge Hamiltonian takes the form}
{%
\begin{equation}
\label{eq:ham_gamma}
\begin{split}
   \hat{H}_\text{L} &= \hat{H}_\text{MBD}+ \sum_{\alpha=1}^{N_p} \Biggl[ \frac{1}{2}\hat{p}_{\alpha}^{2} \\
   &+ \frac{1}{2} \biggl(\omega_{\alpha}\hat{q}_{\alpha}- \sqrt{N_c}\, \boldsymbol{\lambda}_{\alpha} \cdot \hat{\textbf{D}}_{\textbf{q}=0} \biggr)^{2} \Biggr],
\end{split}
\end{equation}}

where {$\hat{\textbf{D}}_{\textbf{q}=0} = \sum_i^{N_a}\sum_a^3 \omega_{ia}\sqrt{\alpha_{ia}}\,\hat\chi_{ia}(\textbf{q}=0)\,\boldsymbol{u}_a$ is the collective dipole operator of the unit cell with Cartesian components $\hat{D}_a$}. The photon mode couples only at one {$\textbf{q}$-point}, i.e.{,} $\textbf q = 0$, which is the $\Gamma$-point. {Per unit cell,} the light-matter coupling is enhanced by {$\sqrt{N_c}$, where $N_c$ is} the number of unit cells.

{For a single $\textbf{q}$-point ($N_c = 1$), Eq.~(\ref{eq:ham_gamma}) reduces to the length gauge pMBD Hamiltonian for finite systems as given in Ref.~\cite{tasci2025},}
\begin{align}
\label{eq:rpa-ham2}
       \hat{H}_\text{L}&=\frac{1}{2}\sum_{i=1}^{N_a}\sum_{a=1}^{3}\left(\pi^{2}_{ia}+\omega^2_{ia}\chi_{ia}^{2}\right)\nonumber\\
       &+\frac{1}{2}\sum_{i,j=1}^{N_a}\sum_{a,b=1}^3\omega_{ia}\omega_{jb}\sqrt{\alpha_{ia}\alpha_{jb}}\chi_{ia}T_{\text{LR},ij}^{ab}(\textbf R_j - \textbf R_i)\chi_{jb}\nonumber\\
       &+\frac{1}{2}\hat{p}_{\alpha}^{2} +\frac{1}{2}(\omega_{\alpha}\hat{q}_{\alpha}-\sum_{i=1}^{N_a}
    \sum_{a=1}^3\lambda_{\alpha a}\omega_{ia}\sqrt{\alpha_{ia}}\chi_{ia})^{2}.
\end{align}

%This Hamiltonian will be used in the following for all numerical simulations.

\section{Numerical details}

{All DFT calculations are performed with the VASP software package~\cite{vasp_PW, vasp_PAW}, using the PBE exchange-correlation functional~\cite{PBE}, the standard PBE PAW potentials for B, N, and C, a plane-wave cutoff of 520 eV, a $\Gamma$-centered $31\times31\times1$ \textbf{k}-point grid, Gaussian smearing with a width of 0.05 eV, and an electronic convergence criterion of $10^{-10}$ eV. The bilayers are constructed from the relaxed monolayer in-plane lattice, with in-plane lattice constants of 2.51~\text{\AA} (hBN) and 2.47~\text{\AA} (graphene), in four-atom unit cells with an out-of-plane cell height of 20~\text{\AA}, where the two layers are placed symmetrically about the cell mid-plane. The binding curves are sampled over interlayer distances from $R = 3.0~\text{\AA}$ to $5.1~\text{\AA}$ in steps of $0.0025~\text{\AA}$, and the reference energy is computed for the same cell with the layers $10~\text{\AA}$ apart. Since the in-plane lattice and the cell height remain fixed, $\Delta E(\lambda_\alpha, R) = E(\lambda_\alpha, R) - E(\lambda_\alpha, 10~\text{\AA})$ is a well-defined binding energy. The per-atom Hirshfeld ratios obtained from the DFT calculation parameterize the polarizabilities and frequencies of the MBD oscillators~\cite{tkatchenko2012accurate}, and we set the MBD damping parameter to the value that is used for the PBE exchange-correlation functional ($\beta=0.83$). The single cavity mode is polarized in the out-of-plane direction with a frequency of $\omega = 2$ eV, and coupling strengths are scanned from $\lambda_\alpha = 0$ to $0.2$ a.u. in steps of $0.02$ a.u. As described in the main text, the cavity-induced shift of {$E_\text{xc}$} is computed on the $5\times3\times1$ $\textbf{q}$-grid (using the effective coupling strength $\lambda_{\alpha,\text{eff}} = \sqrt{N_c}\,\lambda_\alpha$ with $N_c = 15$), while the bare MBD energy is computed on a converged $37\times37\times1$ $\textbf{q}$-grid at $\lambda_\alpha = 0$.

The equilibrium interlayer distance $R_0$, minimum energy, and layer breathing mode (LBM) frequency at each coupling strength are extracted by fitting the binding curve to a third-order polynomial in a window of $\pm 0.0875~\text{\AA}$ (hBN) and $\pm 0.0625~\text{\AA}$ (graphene) around the discrete minimum. $R_0$ is the stationary point of the fit, and the LBM is treated as a rigid-layer breathing mode of two identical layers with reduced mass $\mu = m_\text{layer}/2$, such that $\omega_\text{LBM} = \sqrt{2k/m_\text{layer}}$, where $k$ is the curvature of the fit at $R_0$ and $m_\text{layer}$ is the mass per layer of the unit cell.}

\section{Equivalence Between unit cell and Supercell Calculations}

\begin{figure}[h]
    \centering
    \includegraphics[scale=0.4]{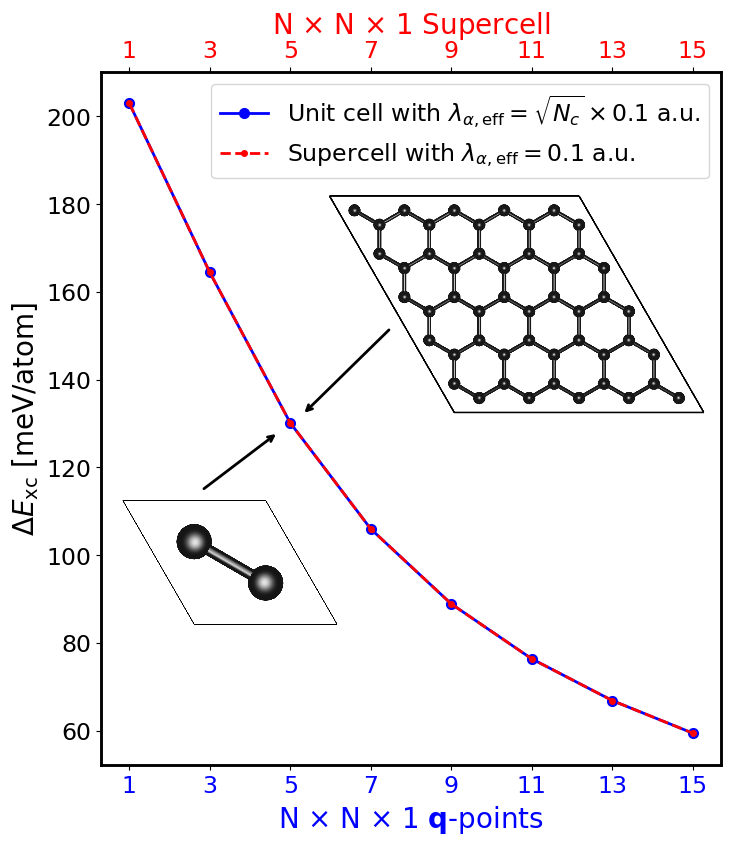}
    \caption{Cavity-modified energies, $\Delta E_\text{xc}$, per atom for unit cell and supercell calculations. The two calculations are equivalent when the number of {$\textbf{q}$-points} along each axis for the unit cell is equivalent to the number of cells along each axis for the supercell, and the coupling strength for the unit cell calculation is scaled by the square root of the total number of {$\textbf{q}$-points}.}
    \label{fig:unit_supercell}
\end{figure}

We first validate our implementation by establishing the equivalence between calculations performed on a {unit cell} with {$\textbf{q}$-point} sampling, and calculations done on a supercell with $\Gamma$-point sampling. In Fig{.}~\ref{fig:unit_supercell}, we show this equivalence for the cavity-mediated energies, $\Delta E_\text{xc}$, of an AA{-}stacked bilayer graphene near the equilibrium interlayer distance ({$R = 3.6425~\text{\AA}$}). Since analysis of the cavity-mediated energy does not require DFT calculations, we fix the Hirshfeld ratios of the carbon atoms for the supercell to those obtained from a DFT calculation on the unit cell. For the MBD calculation, we set the damping parameter to the value that is used for the PBE exchange-correlation functional. The {unit cell} is coupled to a single out-of-plane mode with a coupling strength of 0.1 a.u. and a cavity frequency of 2 eV. Our results show that photon energies in a supercell of size $N_c = N\times N\times 1$ sampled at only the $\Gamma$-point with a coupling strength of $\lambda^{(s)}$ become identical to a simulation on the {unit cell} sampled over an $N_c = N \times N\times 1$ $\textbf{q}$-grid with a coupling strength of $\lambda^{(u)} = \sqrt{N^2}\,\lambda^{(s)} = \sqrt{N_c} \lambda^{(s)}${, which is precisely the effective coupling strength $\lambda_{\alpha,\text{eff}} = \sqrt{N_c}\,\lambda_\alpha$ introduced in the main text}.

\section{Shifts in Interlayer Distances and Vibrational Frequencies in Bilayer Graphene}

\begin{figure}[H]
    \centering
    \includegraphics[scale=0.32]{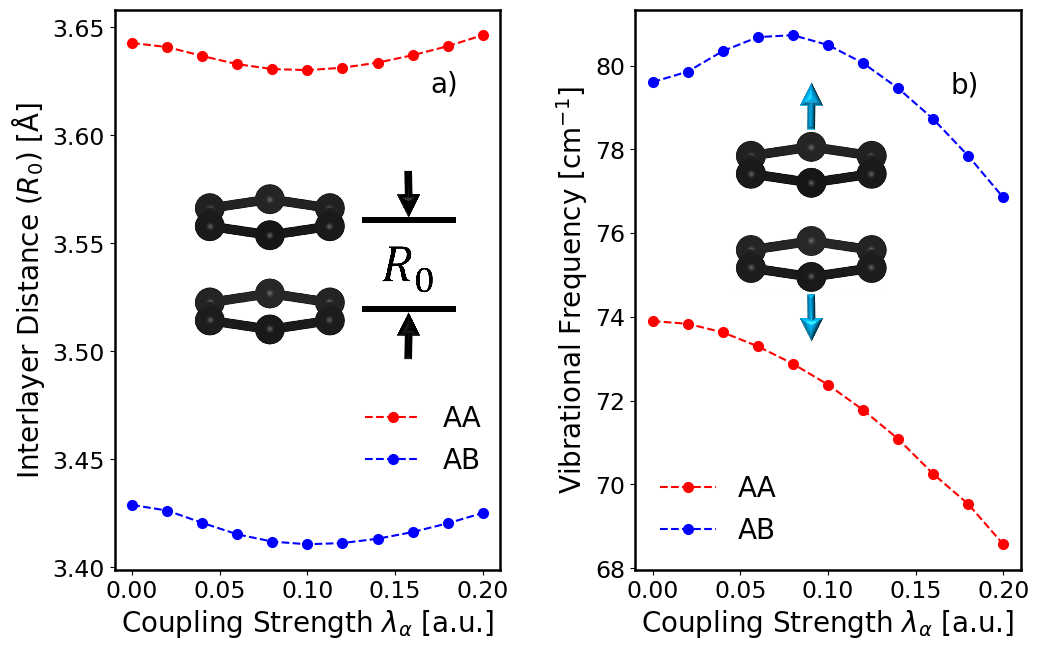}
    \caption{Interlayer distances (a) and layer breathing mode frequencies (b) as a function of coupling strength for bilayer graphene in a cavity with single cavity mode polarized in the out-of-plane direction with a cavity frequency of $\omega = 2$ eV.}
    \label{fig:freq_graphene}
\end{figure}

In Fig.~\ref{fig:freq_graphene}a, we plot the equilibrium interlayer distance with respect to coupling strength for bilayer graphene in the AA and AB stacking configurations. Contrary to what we observe with bilayer hBN, we find that for coupling strengths less than $\sim 0.10$ a.u., the equilibrium interlayer distance decreases slightly before increasing again as the coupling strength increases.
In Fig.~\ref{fig:freq_graphene}b, we plot the LBM frequencies. For AB stacking, the LBM at $\lambda=0$ a.u. (79.6 cm$^{-1}$) is close to what has been measured through Raman spectroscopy~\cite{lui2013breathing}. As the coupling strength increases, the LBM frequency drops by 7.2\% and 3.5\% of the value at $\lambda=0$ a.u. for AA and AB stacking, respectively.

\section{Change in Minimum Energy vs. Coupling Strength}

\begin{figure}[H]
    \centering
    \includegraphics[scale=0.32]{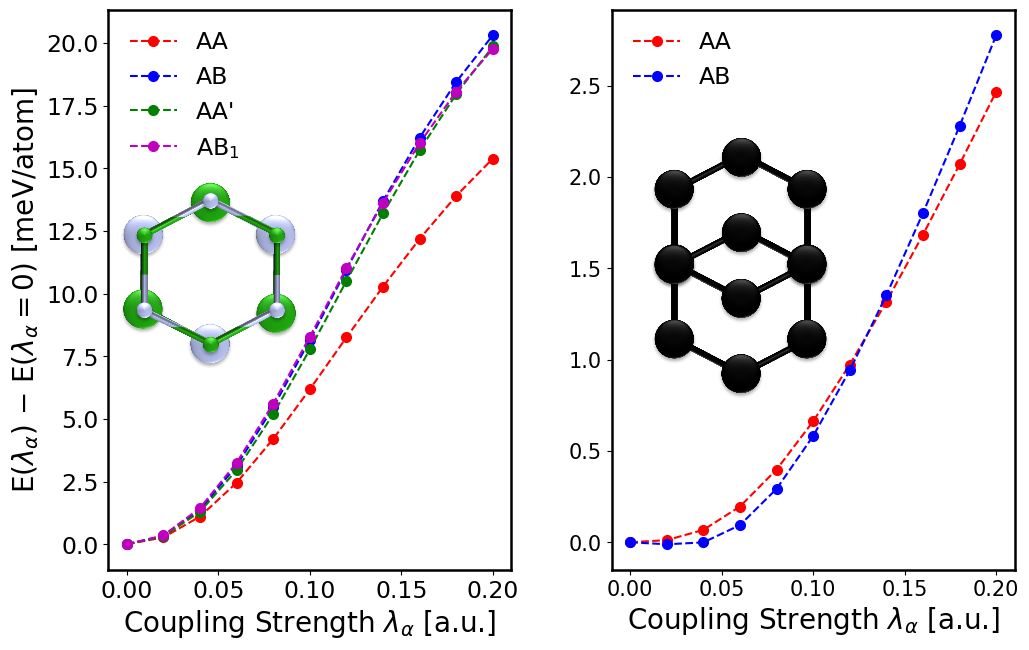}
    \caption{Difference in energy inside and outside of the cavity for different stacking configurations of bilayer hBN (a) and bilayer graphene (b). The energy inside and outside of the cavity is calculated with respect to an offset, in which case $\Delta E(R) = E(R) - E(10~\text{\AA})$. Stacking configurations with higher {polarizabilities} have a faster rate of change with respect to coupling strength.}
    \label{fig:delta_E_min}
\end{figure}

In Fig.~\ref{fig:delta_E_min}, we plot the difference between the equilibrium binding energies inside the cavity and outside of the cavity as a function of coupling strength. In Fig.~\ref{fig:delta_E_min}a, we plot this energy difference for the AA, AB, AB$_1$, and AA' stacking configurations of bilayer hBN, and for the AA and AB stacking configurations of bilayer graphene in {Fig.~\ref{fig:delta_E_min}b}. In general, we observe that stacking configurations for which the binding energies are largest outside the cavity exhibit a faster rate of change in energy with respect to coupling strength. The results in Fig.~\ref{fig:delta_E_min} then imply that changes in preferential stacking order are accessible at lower coupling strengths for stacking configurations {that} are very similar in binding energy or very different in polarizability.

\section{Interpretation of the Coupling Strength}

{
In the main text, the coupling strength $\lambda_\alpha$ and the number of unit cells $N_c$ are treated as independent parameters. In a realistic cavity they may not be independent, because both can be fixed by the electromagnetic environment, which is itself determined by the geometry of the cavity, the boundary conditions, the dielectric response, and the cavity losses~\cite{fan2026unified}. For a given cavity design, electromagnetic simulations yield the coupling strength directly~\cite{svendsen2024general}, and the corresponding effective mode volume follows by inverting $\lambda_\alpha = \sqrt{4\pi/V_\alpha}$.

Writing the mode volume as $V_\alpha = A_\alpha L_\alpha$, where $A_\alpha$ is the mode area and $L_\alpha$ is the characteristic length of the cavity, makes the constraint explicit. When a 2D material couples to the out-of-plane polarization, it fills the entire extent of the mode area, so the number of unit cells that couple to the cavity mode is set by the mode area rather than chosen freely, $N_c = A_\alpha/A_\text{uc}$, where $A_\text{uc}$ is the area of the unit cell. The effective coupling strength introduced in the main text then becomes
$\lambda_{\alpha,\text{eff}} = \sqrt{\frac{4\pi}{A_\text{uc}L_\alpha}}$,
in which the mode area cancels. Enlarging $A_\alpha$ admits more unit cells into the coherently coupled region and enhances the coupling by $\sqrt{N_c}$, but it dilutes the field by the same factor, so that $\lambda_{\alpha,\text{eff}}$ depends only on the unit cell area and the cavity length.

For a given material and cavity frequency, the cavity enters the $\Gamma$-point energy only through $\lambda_{\alpha,\text{eff}}$, so the $\Gamma$-point contribution is independent of $A_\alpha$ at fixed $L_\alpha$. The cavity-mediated energy defined in the main text is this contribution normalized by the number of sampled \textbf{q}-points. Since the \textbf{q}-point grid samples exactly those cells that couple to the mode (assuming $N_q = N_c$), we therefore find 
$\Delta E_\text{xc} \propto {1}/{A_\alpha}$.
Thus, at fixed cavity length, the cavity-mediated energy per unit cell is inversely proportional to the mode area.
}

\begin{figure}[H]
    \centering
    \includegraphics[scale=0.45]{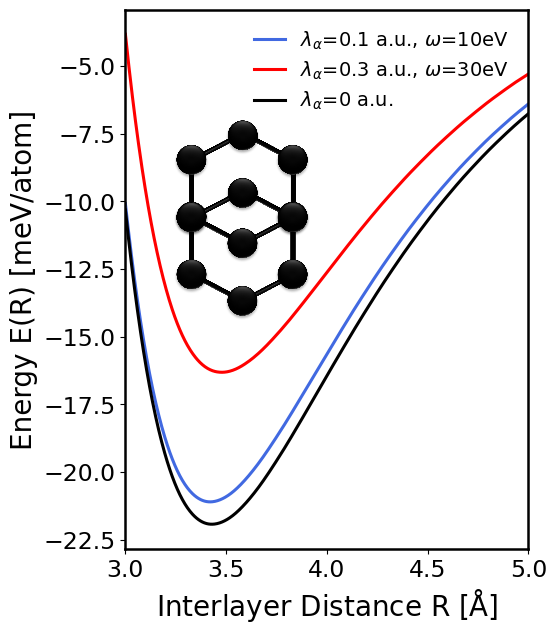}
    \caption{Binding curves for bilayer graphene in AB stacking for the same ${\lambda_\alpha}/{\omega}$ ratio.}
    \label{fig:lambda_omega_ratio}
\end{figure}

{ 
Recent work on cavity-modified van der Waals materials shows that binding energies depend only on the ratio of the coupling strength to the cavity frequency ($\lambda_\alpha / \omega_\alpha$)~\cite{I-Te_epLDA, Liu_QEDFTvdW}. In Fig.~\ref{fig:lambda_omega_ratio}, we analyze this dependence for the pMBD functional and find that the binding energies are not solely dependent on this ratio. The binding energies depend separately on the coupling strength and cavity frequency, due to correlation effects (e.g.  higher-order photon processes) captured by the pMBD framework.}

% \textcolor{purple}{Recent work on cavity-modified van der Waals materials quantifies coupling strength with the ratio $\frac{\lambda_\alpha}{\omega}$~\cite{I-Te_epLDA, Liu_QEDFTvdW}. This is possible because the QEDFT formulation used there relies on $\lambda_\alpha$ only as part of the $\frac{\lambda_\alpha}{\omega}$ ratio. In Fig.~\ref{fig:lambda_omega_ratio}, we analyze this dependence and find that the binding energies are not solely dependent on this ratio. The binding energies are depend separately on the coupling strength and cavity frequency, owing to the higher order photon processes captured by the pMBD framework.}

\bibliography{refs} % Produces the bibliography via BibTex.